\documentstyle[psfig,12pt]{article}

\font\cmr=cmr7

\newcommand{\be}{\begin{equation}}
\newcommand{\ee}{\end{equation}}
\newcommand{\bea}{\begin{eqnarray}}
\newcommand{\ena}{\end{eqnarray}}

\newcommand{\slP}{\raise.15ex\hbox{$/$}\kern-.53em\hbox{$P$}}
\newcommand{\slR}{\raise.15ex\hbox{$/$}\kern-.53em\hbox{$R$}}
\newcommand{\slL}{\raise.15ex\hbox{$/$}\kern-.53em\hbox{$L$}}

\newcommand{\dsiggg}{ {d \sigma^{\gamma \gamma}
            \over {d {\vec p}_T d \eta}}}
\newcommand{\dsiggq}{{d \sigma^{\gamma q}
            \over {d {\vec p}_T d \eta}}}
\newcommand{\dsigd}{  {d \sigma^{D} \over {d {\vec p}_T d \eta}}}
\newcommand{\dsigsf}{{d \sigma^{SR} \over {d {\vec p}_T d \eta}}}
\newcommand{\dsigdf}{{d \sigma^{DR} \over {d {\vec p}_T d \eta}}}

\def\pT{p_{_T}}
\def\pqgam{P_{q/\gamma}}

\def\alf{\alpha}

\def\alfs{\alpha_s}
\def\alfspi{{\alpha_{s} \over 2 \pi}}

\def\be{\begin{equation}}
\def\ee{\end{equation}}
\def\bea{\begin{eqnarray}}
\def\eea{\end{eqnarray}}

\def\Journal#1#2#3#4{{#1} {\bf #2}, #3 (#4)}

\def\NPB{{\em Nucl. Phys.} B}
\def\PLB{{\em Phys. Lett.}  B}
\def\PRL{\em Phys. Rev. Lett.}
\def\PRD{{\em Phys. Rev.} D}
\def\ZPC{{\em Z. Phys.} C}

\begin{document}
\renewcommand{\thefootnote}{\fnsymbol{footnote}}
\newpage
\pagestyle{empty}
\setcounter{page}{0}

\def\logoenslapp{\logolight}
%
%
%
%
\newcommand{\norm}[1]{{\protect\normalsize{#1}}}
\newcommand{\LAP}
{{\small E}\norm{N}{\large S}{\Large L}{\large A}\norm{P}{\small P}}
\newcommand{\sLAP}{{\scriptsize E}{\footnotesize{N}}{\small S}{\norm L}$
${\small A}{\footnotesize{P}}{\scriptsize P}}
\def\logolapin{
  \raisebox{-1.2cm}{\epsfbox{/lapphp8/keklapp/ragoucy/paper/enslapp.ps}}}
\def\logolight{{\bf{{\large E}{\Large N}{\LARGE S}{\huge L}{\LARGE
        A}{\Large P}{\large P}} }}
\begin{minipage}{5.2cm}
  \begin{center}
    {\bf Groupe d'Annecy\\ \ \\
      Laboratoire d'Annecy-le-Vieux de Physique des Particules}
  \end{center}
\end{minipage}
\hfill
\logoenslapp
\hfill
\begin{minipage}{4.2cm}
  \begin{center}
    {\bf Groupe de Lyon\\ \ \\
      Ecole Normale Sup\'erieure de Lyon}
  \end{center}
\end{minipage}
\\[.3cm]
\centerline{\rule{12cm}{.42mm}}

\vspace{20mm}

\begin{center}

{\LARGE {\bf  Inclusive hard processes in photon induced reactions
}\footnote{Talk
presented at PHOTON '97, Egmond aan Zee, The Netherlands, 
May 1997}}\\[1cm]

\vspace{10mm}

{\large P.~Aurenche$^{1}$}\\[.42cm]

{\em Laboratoire de Physique Th\'eorique }\LAP\footnote{URA 14-36 
du CNRS, associ\'ee \`a l'Ecole Normale Sup\'erieure de Lyon et
\`a l'Universit\'e de Savoie.}\\[.242cm]

$^{1}$ Groupe d'Annecy: LAPP, BP 110, F-74941
Annecy-le-Vieux Cedex, France.


\end{center}
\vspace{20mm}

\centerline{ {\bf Abstract}}

\indent
In the following some aspects of inclusive hard processes in photon
induced reactions are reviewed. After a discussion on the properties
of hard processes, the phenomenology of jet production and
of charmonium production is presented in the context of the
next-to-leading logarithm approximation of QCD.

\vfill
\rightline{hep-ph/yymmnn}
\rightline{\LAP-A-652/97}
\rightline{May 1997}

\newpage
\pagestyle{plain}
\renewcommand{\thefootnote}{\arabic{footnote}}

\section{Introduction}

Over the last few years the physics of hard processes in photon-induced
reactions ({\em e.g.} photo-production at HERA and photon-photon
collisions at TRISTAN and LEP) has almost reached the status of
sophistication of hard processes in purely hadronic reactions. The
experimental results are becoming more and more accurate and many
observables measured in hadronic reactions are now being accessible in
photon-induced reactions while the corresponding theoretical
calculations have been performed at the next-to-leading order (NLO) of
perturbative QCD. Since PHOTON~'95 \cite{photon} an enormous progress
has been made in this respect so that one should be able to start
quantitative phenomenology. The main points that deserve special
attention now are: 1) matching the calculated observables with the
experimental ones; 2) matching the appropriate non-perturbative input to
the conventions used in the NLO calculations. 

It is well known that the physics of photon-induced processes is more
complex than that of pure hadronic reactions. This arises from the fact
that the photon acts either as a {\em parton} which couples directly to
the hard scattering and contributes to $direct$ processes, or  
as a $composite$ object (a bag of partons) the constituents
of which couple to the hard scattering via the photon structure
functions: this leads to $resolved$ processes. Furthermore, as discussed
in  ref. \cite{vogt} the photon structure function contains two pieces:
\bea
F_{i/\gamma}(x,Q) =  F^{\mbox{\cmr had}}_{i/\gamma}(x,Q) \  
+\ F^{\mbox{\cmr anom}}_{i/\gamma}(x,Q),\qquad  i={\mbox {quark, gluon}}
\ena
where the first term on the right hand-side is similar to the hadronic
structure function while the second term increases logarithmically 
with $Q^2$ and is asymptotically calculable in perturbation theory.

In the following, in a theoretical introduction, the structure of a
``typical" cross section will be derived stressing the  features
distinguishing the $direct$ processes from the $resolved$ ones. Two
topics  are selected for further discussion: the production of jets and
single particle on the one hand and the production of hidden charm for
which  HERA offers the possibility to test models of $J/\Psi$ production
at the Tevatron.

\section{Photon induced reactions: theory}\label{sec:theory} 
\begin{figure}[h]
\vspace{-.1cm}
\centerline{\psfig{figure=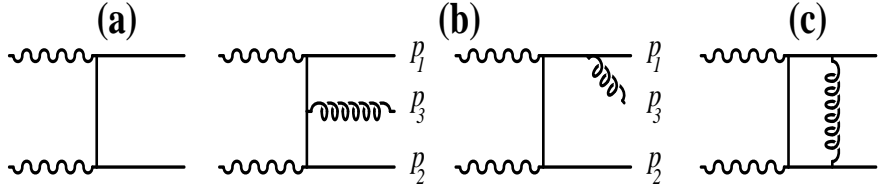}}
\caption{{\em Feynman diagrams for $\gamma\gamma$ scattering. 
({\bf a}): Born approximation;\ ({\bf b}) real corrections;
\ ({\bf c}) virtual corrections.}} 
\label{fig:1}
\vspace{-.1cm}
\end{figure}

\begin{figure}[h]
\vspace{-.1cm}
\centerline{\psfig{figure=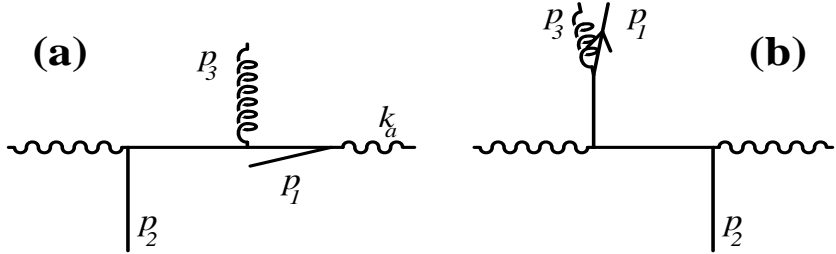}}
\vspace{-.3cm}
\caption{{\em Diagrammatic representation of the collinear singularities.}}
\label{fig:2}
\vspace{-.1cm}
\end{figure}

We consider photon-photon collisions as an example. The lowest order
diagram is shown in fig. \ref{fig:1}a. To apply the perturbative approach
a large scale is needed which is provided by the transverse
momentum of the produced jet or the mass of the  heavy quark.
Considering jet production, the Born approximation is valid for large
values of $\pT/\sqrt s$  where unfortunately the cross section is very
small since it behaves as $d\sigma^{\cmr jet} / d{\vec p}_T \sim \alf^2
/p^{4}_T .$ To obtain reliable predictions at lower $\pT$ values it is
necessary to consider diagrams with an extra gluon emission, leading in
principle, to ${\cal O}(\alfs)$ corrections to the cross section  (fig.
\ref{fig:1}b,c). When integrating over the phase-space of final state
partons to reconstruct {\em e.g.} the single inclusive jet cross
section, one encounters ``dangerous" regions where the virtualities of
some fermion propagators may vanish and the corresponding matrix
elements are not defined. This happens, for example, when the final
state parton of momentum $p_{_1}$ is collinear to the initial photon of
momentum $k_a$ ($s_{a_1}=(k_a-p_{_1})^2 \rightarrow 0$) or the emitted
gluon is collinear to a final state quark ($s_{_{13}}=(p_{_1}+p_{_3})^2
\rightarrow 0$). 
Considering the $s_{a_1}$ singular
region, the process can be pictured as in fig. \ref{fig:2}a where
the initial photon fragments into a collinear $q \bar q$ pair  followed
by the hard 2 body $\rightarrow$ 2 body scattering of the $q$ or $\bar
q$, the other parton flying off down the beam pipe. The photon fragmentation
is a soft process (the relevant scale $s_{a_1}$ is small) which takes
place on a long time scale before the short-distance process scatters
the partons at large transverse momenta. The cross section can be written as 
\be
\dsiggg  =   \int dz \pqgam (z)
\ln\left(M^2 \over \lambda^2\right)\dsiggq \ + \ {\alfspi} K^{D}(\pT,M),
\label{eq:perturb}
\ee
where $\lambda$ is a cut-off introduced to regularise the collinear
singularity, and $M \sim {\cal O}(\pT)$ is an arbitrary scale introduced
to separate the  contribution from the collinear region, generating the
``large" $\ln(M^2 /\lambda^2)$ piece, from the hard region (rouhgly
$s_{a_1}>M^2$) leading to the term $K^D$ containing only finite terms
such as $\ln(\pT /M^2)$.  The above expression is rigourously
independent of the scale $M$.  In fact, at this level of approximation
the scale $M$ is not necessary  but it is introduced for later use. The
integration variable in eq. \ref{eq:perturb} is simply the fraction of
the photon momentum carried by the interacting $q$ or $\bar q$. Using
the factorisation theorem, we can substitute to the ``large logarithm"
term $ \pqgam (z) \ln(M^2 / \lambda^2)$ the photon structure function
$F_{q/\gamma}(z,M)$, measured in $\gamma^* \gamma$  collisions, which
satisfies evolution equations of type:
\be
{d F_{i/\gamma}(M) \over d \ln M^2} =   P_{i \gamma}
+ \sum_{j=q,g}P_{ij} \otimes F_{j/\gamma}(M), \qquad i,j=q,g
\label{eq:evolsimp}
\ee
where the symbol $\otimes$ denotes a convolution over the longitudinal
variable and the $P_{ij}$ are the usual splitting functions. After
replacing $ \pqgam \ln(M^2 / \lambda^2)$ by $F_{q/\gamma}(M)$
in eq. (\ref{eq:perturb}) the compensation in the $M$ dependence on the
right-hand side is only approximate as the evolution equation
effectively resums large higher order corrections (of type $[\alpha_s(M)
\ln(M / \Lambda_{_{QCD}})]^n$) in $F_{q/\gamma}(M)$ whereas no such
resummation is performed in $K^D$. The cross section can be written
\bea
\dsiggg = \left( \dsiggg^{(0)} + {\alfspi} K^D(\pT,M) \right) +
\int dz F_{q/\gamma}(z,M) \dsiggq 
\label{eq:sr}
\ena
where the $K^D$ term is a correction to the Born cross section
$d\sigma^{\gamma\gamma(0)}$ and  contributes to the
{\em direct} cross section evaluated in the NLO approximation. The last
term in the equation is the {\em resolved} process. Since in the latter
case only part of the energy of the photon contributes to the
large $\pT$ process, the rest being carried  by the longitudinal
fragment, the $resolved$ term is qualitatively 
({\em i.e.} in the LO approximation) distinguished from the
$direct$ one by: \\ 
- a softer $\pT$ spectrum; \\ 
- a jet system boosted in the direction opposite to that of the
resolved photon: {\em e.g.} a backward moving resolved
photon produces forward going jets; \\
- some hadronic activity  in the direction of the resolved photon. \\ 
In the NLO approximation the {\em direct} and {\em resolved} components
are {\em not separatly observable} since they depend on the arbitrary
renormalisation scale $M$. However, using the properties above it is
possible to define observables related to these two components (see
below).

Until now we have discussed features related to the initial state
singularities. The final state singularities (of type $s_{_{13}}
\rightarrow 0$) also lead to 2 body $\rightarrow$ 2 body hard
scattering with the singular behaviour associated to the fragmentation
process exactly as in purely hadronic processes (fig. \ref{fig:2}b). The
cross section is made finite by adding the virtual diagrams and properly
defining the jets, merging the two almost collinear partons into one
jet, or by convoluting with a fragmentation function and using the
factorisation theorem to build the scaling violations in the
fragmentation function.

One could continue the perturbative analysis and consider the emission
of more partons: this will lead to {\em double resolved} processes where
each photon interacts via its structure function. Finally, the
$\gamma\gamma$ cross section for inclusive jet production takes the
form: 
\be
\frac{d\sigma^{\gamma\gamma}}{d\vec{p}_T d \eta} \ = \ \dsigd + \
\dsigsf + \ \dsigdf 
\label{eq:sum}
\ee
with the $direct$, $single\ resolved$ and $double\ resolved$ pieces given in
the NLO approximation by an expansion of type:
\be
\dsigd (R) =  \dsiggg^{(0)} \ + \ \alfspi K^{D} (R;\pT ,M),
\label{eq:dir}
\ee
and similarly for the other terms.
The variable $R$ specifies the jet cone size. The factorisation scale
variation associated to the inhomogeneous
term $P_{i \gamma}$ in eq. (\ref{eq:evolsimp}) is compensated between
the $D$, $SR$ and  $DR$ pieces while the variation associated to the
homogeneous terms $P_{ij}$ compensates within each of these pieces
between the lowest order and the correction terms. Thus only eq.
(\ref{eq:sum}) is expected to be stable under variation of scale $M$ but
not its individual components. \\
For hadron production an extra convolution of the above expresssions
with the relevant fragmentation function is needed. For photo-production
processes one of the $F_{i/\gamma}(x,M)$ should be replaced by the
hadronic structure function. In an obvious notation, the cross section
takes the simpler form:
\be
\frac{d\sigma^{\gamma p}}{d\vec{p}_T d \eta} \ = 
 \frac{d\sigma^{D}}{d\vec{p}_T d \eta} \ + \ 
\frac{d\sigma^{R}}{d\vec{p}_T d \eta}.  
\label{eq:phopro}
\ee

\vskip -0.5cm

\section{Phenomenology of jet production}\label{sec:jet} 

\begin{figure}[t]
\vspace{-.1cm}
\centerline{\psfig{figure=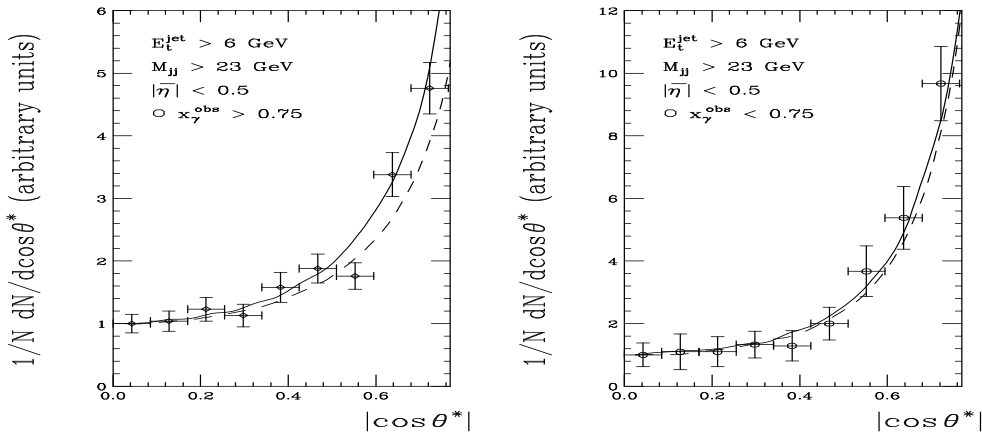,height=6.0cm,width=13.0cm}}
\vspace{-.3cm}
\caption{{\em
Dijet angular distribution as measured by ZEUS \protect \cite{zeusang}
normalized to one at $\cos\theta^*=0$ compared  with LO (dash lines) and
NLO result (solid lines), from ref. \protect \cite{harriso}}.}
\label{fig:4}
\vspace{-.1cm}
\end{figure}

\begin{figure}[t]
\vspace{-.5cm}
\centerline{\hspace{.7cm}\psfig{figure=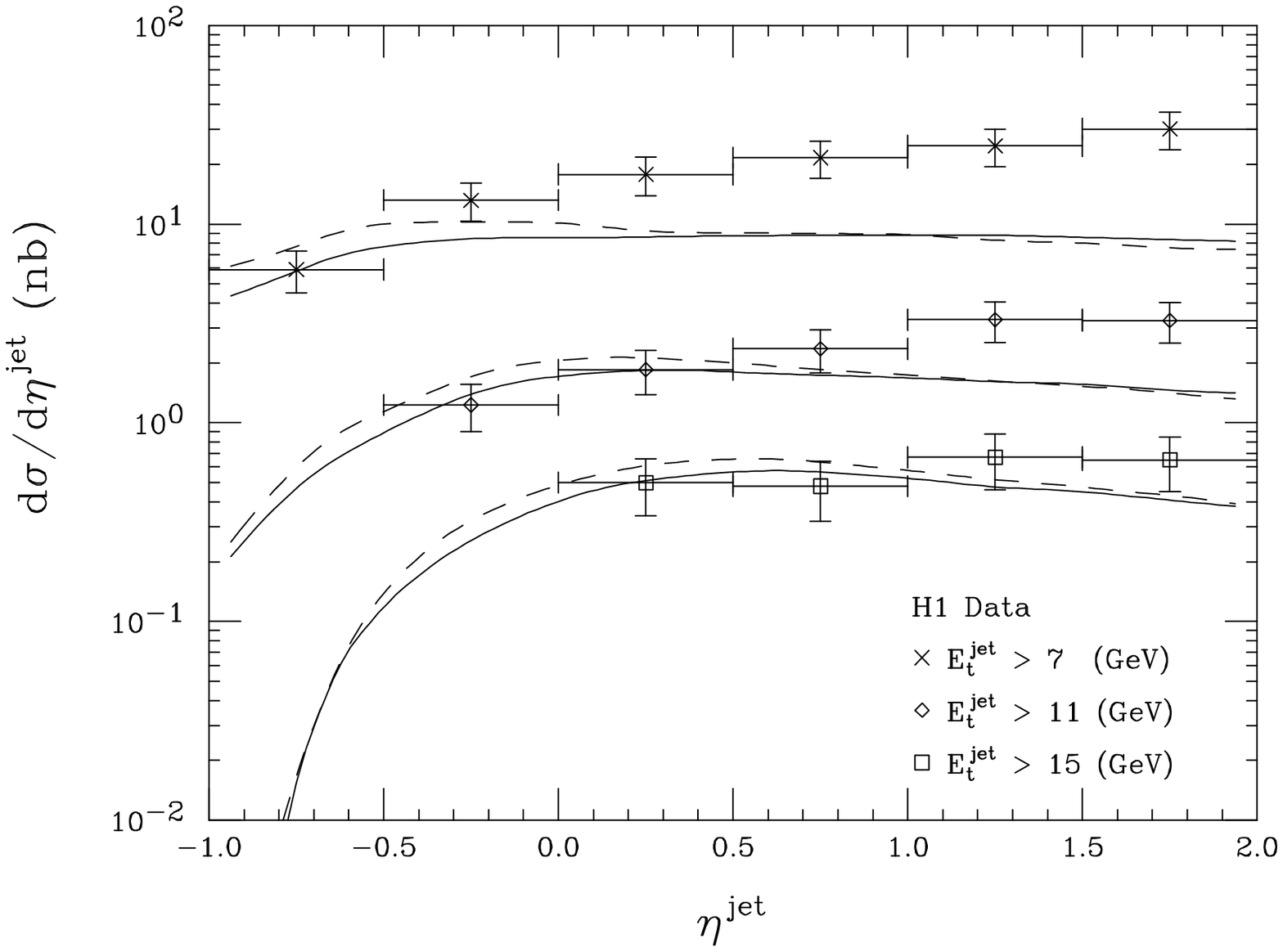,height=6.5cm}
	   \hspace{1.cm}\psfig{figure=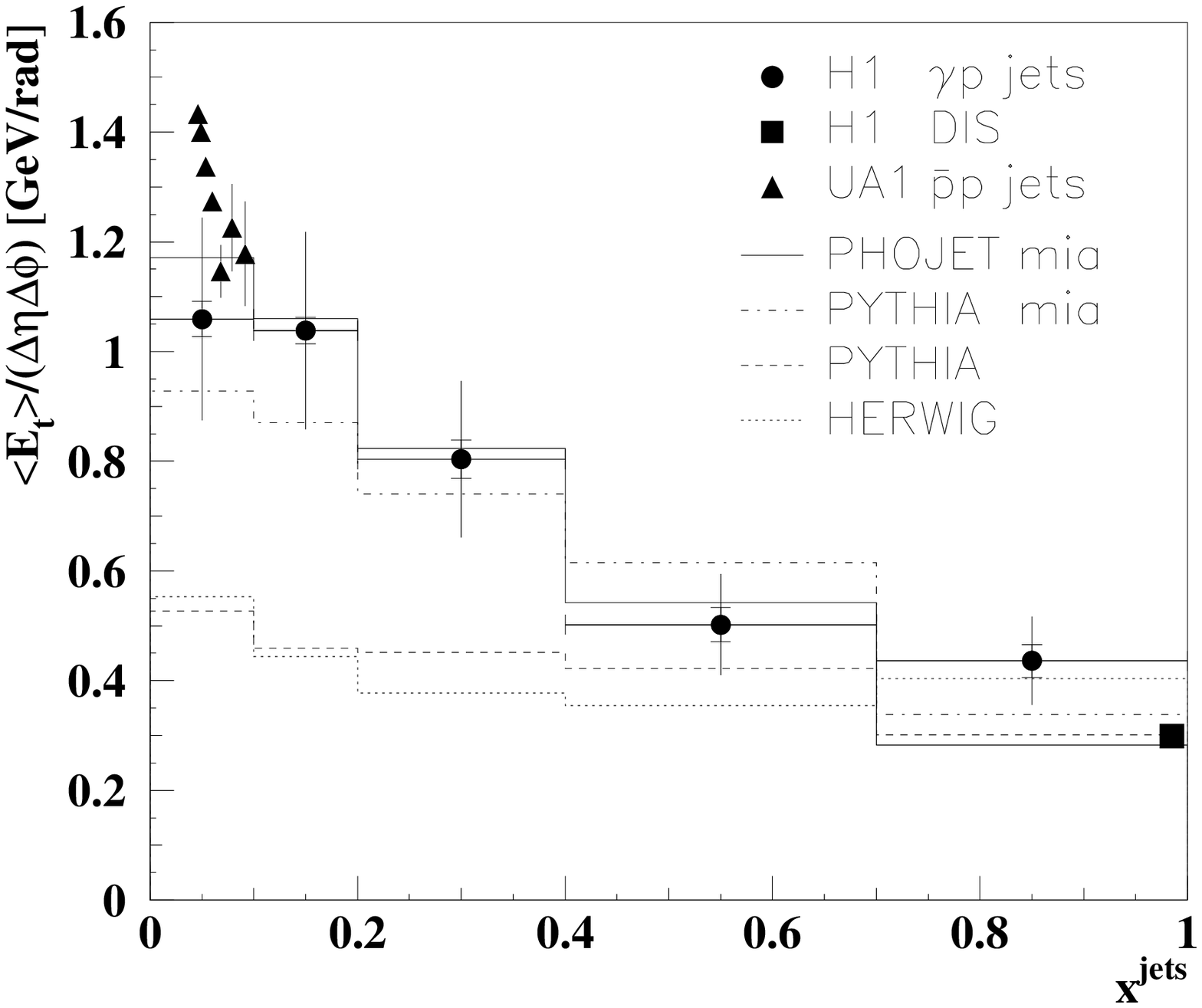,height=7.9cm
						   ,width=8.5cm}}
\vspace{-.8cm}
\caption{{\em
Comparison of the single jet inclusive cross section measured by
H1 \protect\cite{h1jet} compared with NLO
predictions \protect \cite{harriso} using 2 different sets of photon
structure functions.
Transverse energy of the underlying event at HERA as measured by 
H1 \protect \cite{h1jet}.}}
\label{fig:6}
\vspace{-0.1cm}
\end{figure}

One of the aims of studies of hard processes in $\gamma$ induced
reactions is the NLO determination of the photon structure 
function \cite{erdmann}. 
To achieve this it is very useful to isolate in an
experimental way the {\em resolved} component of the cross section. In a
{\em direct} process, all the photon energy is given to jets at large
$\pT$ so that by momentum conservation one has  for a 2-jet event (the
photon is assumed to move towards negative rapidity):
\be 
\vspace{-.3cm}
x_\gamma = \frac{E_{T_1} e^{-\eta_1}+E_{T_2} e^{-\eta_2}}
{2E_\gamma}\ =\ 1 
\ee
Higher order correction are not expected to change this relation
drastically. In contrast, for a {\em resolved} event $x_\gamma < 1$
since part of the photon energy disappears in the beam pipe (fig.
\ref{fig:2}a) (in ref. \cite{bourh} a modified definition of
$x_\gamma$ is proposed). At HERA, the experimental groups choose the value $x_\gamma
= .75$ as a cut-off so that above this value the events are produced
mostly by the {\em direct} process while below they are mainly {\em
resolved}. 
\begin{figure}[htb]
  \centerline{\hspace{-1.cm}\psfig{figure=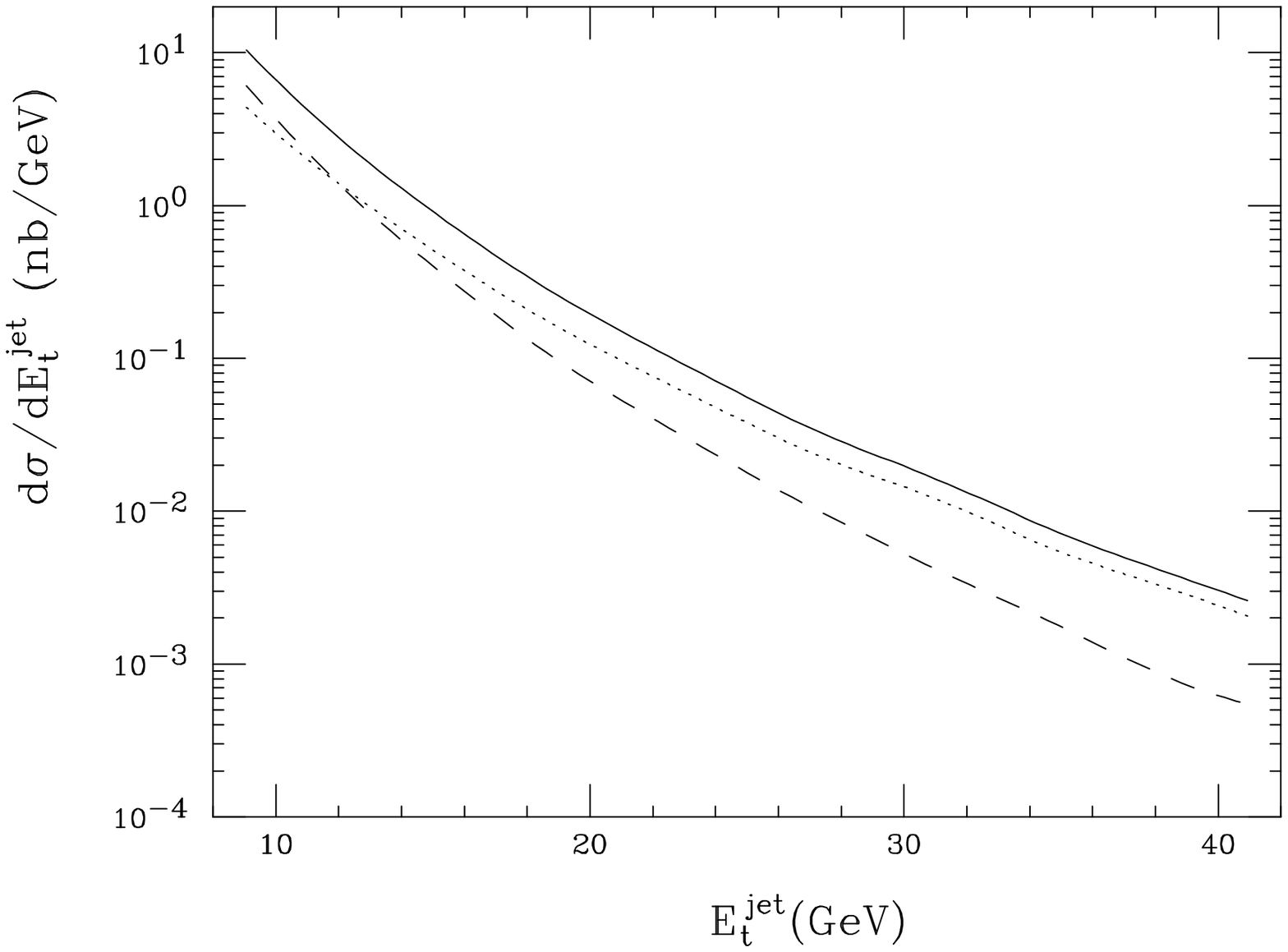,height=6.cm}
  \hspace{1.cm}\psfig{figure=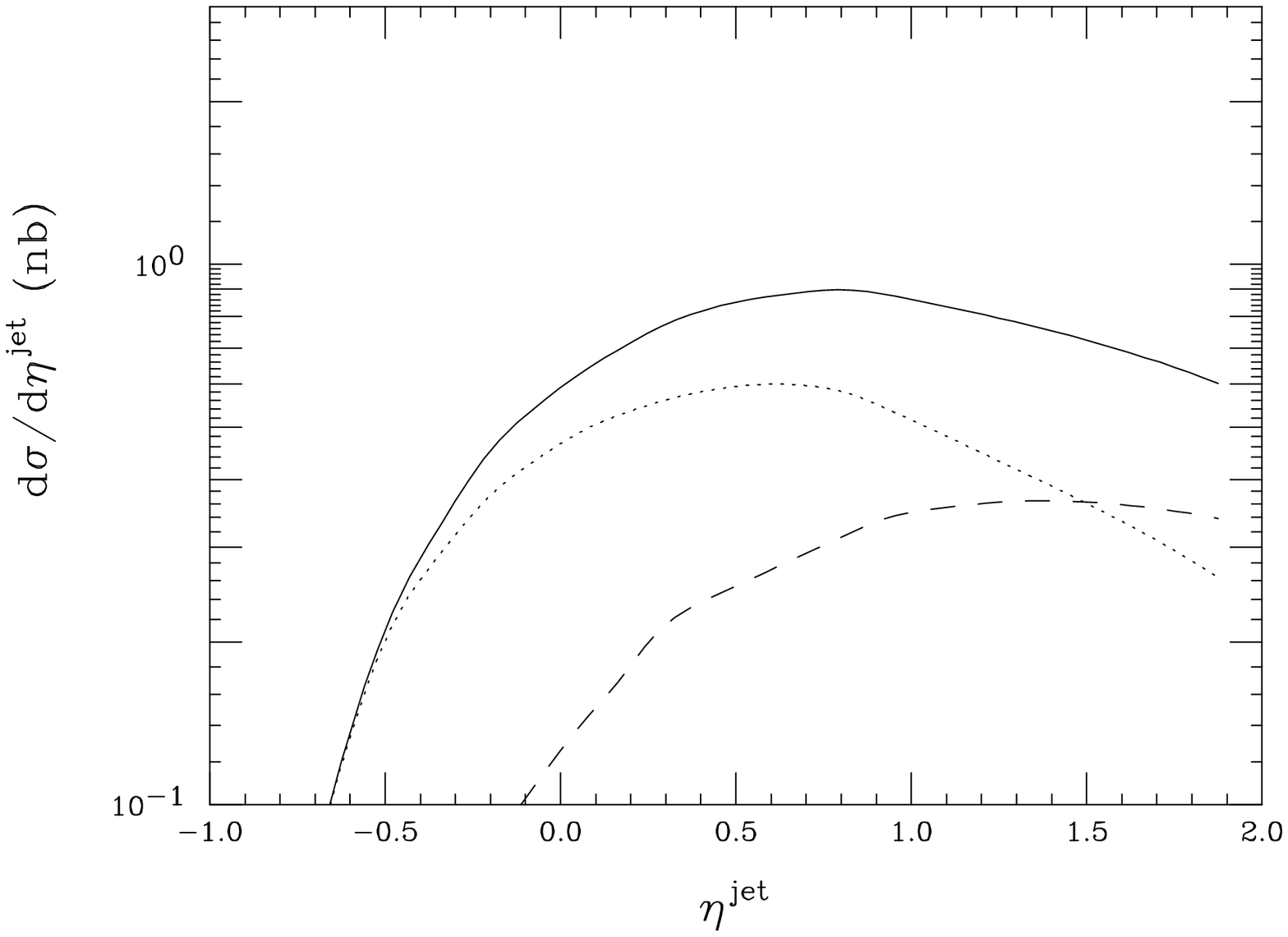,height=6.cm}\hspace{-1.cm}}
  \vspace{-0.3cm}
  \caption[]{\em
Integrated di-jet cross section in the domains $0<x_\gamma<1$ (solid
line), $x_\gamma < .75$ (dashed line), $x_\gamma > .75$ (dotted line).
From ref. \cite{harriso}.}
\label{fig:3}
\vspace{-.1cm}
\end{figure}
The integrated di-jet cross sections for HERA, shown in
fig. \ref{fig:3}, nicely illustrate, in the NLO
approximation \cite{harriso}  the different characteristics of the two
components. In the first figure the direct component shows a faster
decrease with $E_T$, while in the second one the resolved cross section
is seen to contribute mainly forward jets. In fig. \ref{fig:4} a very
good agreement is seen, between the NLO theory \cite{harriso} and
experiment \cite{zeusang}, in the shape of the di-jet angular distribution 
(the angle is measured in the di-jet rest frame): as expected the {\em
resolved} component has a steeper angular dependence due to the
importance of gluon exchange diagrams while for the {\em direct} term
only quark exchange is allowed. 
In the next figure, we compare theoretical predictions of two
independent groups \cite{harriso,klakra} with ZEUS 
data for the di-jet cross section \cite{zeusdij}
$d\sigma / d\bar\eta$, integrated over the phase space $E_{T_1},E_{T_2}
\ge E_0$ and $|\eta_1 - \eta_2| < .5$, with $\bar\eta = (\eta_1 +
\eta_2)/2$. 
The first point to notice is the disagreement between the two sets of
theoretical predictions: this arises because the groups do not calculate
the same observable due to instabilities in the perturbatively
calculated quantity. This is related to the fact that the boundary
condition $E_{T_1}=E_{T_2} = E_0$ is an infrared singular point. 
Typically, in the NLO approximation one needs to consider  the
\begin{figure}[htl]
\vspace{-.1cm}
\vspace{8.5cm}
\caption{{\em 
Dijet cross section vs. ${\overline \eta}$ integrated over  $E_T^{{\rm
jet}}>E_0$ for $E_0=6,8,11,15\, {\rm GeV}$; the four figs. on the left
are as for $x_{\gamma}>0.75$; those on the right are for
$x_{\gamma}<0.75$. The data are from \protect \cite{zeusdij}. Thin lines
from ref. \protect \cite{harriso}; thick dashed lines
from ref. \protect \cite{klakra}}.}
\label{fig:5}
\vspace{-.1cm}
\end{figure}
production of 3 partons in the final state: it can be separated into two
classes:
\bea
\vspace{-0.5cm}
\sigma^{2 \rightarrow3} = \sigma^{2\rightarrow 3}(y_c) + 
\sigma^{2\rightarrow 2}_R(y_c)
\vspace{-.5cm}
\ena
with the parameter $y_c$ such that the first term on the right hand side
contains events generated with ``dangerous" invariant masses 
[{\em i.e.} leading to soft and collinear singularities in the matrix element]
$s_{ij} > y_c s_{_{ab}}$, while $\sigma^{2\rightarrow 2}_R(y_c)$
contains events which look like 2-body hard scattering as discussed
above. This part contains all divergences and $y_c$ is chosen small
enough  ($y_c= 10^{-5}$ to $10^{-2}$) so that approximations can be used
to extract divergences analytically. Upon adding the  virtual
corrections which are also of type 2\ body $\rightarrow$ 2\ body
scattering all divergences cancel and the theoretical cross section is:
\bea
\vspace{-0.5cm}
\sigma = \sigma^{2\rightarrow 3}(y_c) + \sigma^{2\rightarrow 2}(y_c)
\vspace{-.5cm}
\ena
each piece being regular, the first one beeing calculated numerically
and the second one semi-analytically. By histogramming one reconstructs
any observable but it should be checked that the result is independent
of $y_c$. It turns out that for the dijet observable of fig.
\ref{fig:5}, the condition $E_{T_1},E_{T_2} > E_0$ introduces
constraints on the phase space which spoil the $y_c$ compensation
between  the terms in the above equation. Harris and Owens \cite{harriso}
(who use a more elaborate method than the one described above) observe
that the remaining dependence on $y_c$ is much less than the
experimental error bars and present results with some value of $y_c$
while Klasen and Kramer \cite{klakra} modify the boundary condition to
$E_{T_1} > E_0$, allowing for a smaller $E_{T_2}$ if the third
unobserved jet has transverse energy less than $E_{T_3}<1$ GeV. This is
sufficient to remove the $y_c$ dependence but the result is rather
sensitive to the latter energy cut. As seen in the figure this slight
modification of the boundary condition considerably affects the
predictions \cite{harrisow}. If one considers that the calculation of
ref. \cite{harriso} is closer to the experimental observable there
remains a drastic disagreement betwen the data and the theoretical
predictions  specially concerning the $resolved$ cross section.
Disagreement between 
theory \cite{harriso,afg,bodeks} 
and experiment \cite{h1jet} is also seen when comparing
the rapidity distribution of single inclusive jet production as
displayed in fig. \ref{fig:6}: this is particularly marked at low $E_T$
and large pseudo-rapidities, {\em i.e.} where the resolved component
plays a dominant role. This leads us to the important question of
whether or not the theoretical predictions are able to match the
experimental observables.  Two questions can be raised:\\
- is it possible to match theoretical jets, made up of partons, with
experimental jets reconstructed, via various algorithms, from energy
deposited in calorimeter cells? \\
- how does one take into account the transverse energy of the
{\em underlying event}, {\em i.e.} the energy generated by the 
interaction of the {\em spectator} partons?\\
The first point was studied many years ago in connection to jet
production at the Tevatron \cite{ellisks} and was recently re-analysed
in detail for HERA by Kramer and collaborators \cite{butterfkk}: the
idea is to introduce an extra parameter $R_{\mbox{\cmr sep}}$ controling the
width of the jets which is adjusted to fit the data: a better
description of the data  by the NLO calculations is indeed achieved at
the expense, however, of the predictive power of the theory. \\
Concerning the second point the problem can be qualitatively  understood
in the following way. 
At HERA, in events with {\em direct} photo-production of jets the
transverse energy of the {\em underlying event} is produced by
string-like effects: typically a transverse energy of $300-400$ MeV per
unit of rapidity is expected. On the contrary, in resolved
photo-production, the remnants of the photon can undergo soft or semi-hard
interactions with those of the  proton \cite{engelr} generating
transverse energies in the GeV range: the larger the energy of the
photon remnant ($i.e.$ the smaller $x_\gamma$ is) the larger is the
underlying transverse energy: this is illustrated in fig. \ref{fig:6}
from H1 \cite{h1jet} which shows the $E_T$ outside the jets as a function
of $x_\gamma$.  Such an effect is of course present in purely hadronic
reactions but is relatively less important because the much higher $E_T$
values of jets probed in hadronic colliders.
On the other hand, jets in $\gamma\gamma$ reactions should be less
affected except at very low $E_T$ values ($E_T < 5$ GeV at LEP1) where
the {\em double-resolved} process is dominant. Good agreement of
TRISTAN \cite{amy,topaz} and LEP \cite{opal} data with the NLO
predictions \cite{kleinkra} has been obtained for inclusive two-jet
production.\\
\indent 
A discussion about the determination of the gluon density  in the photon
using jet cross sections is given, in the LO approximation, in
ref. \cite{erdmann}.\\
\indent
A way to avoid problems related to jet definition or to underlying
transverse energy is to consider single hadron production. Good
agreement is indeed found in photo-production between the NLO
predictions \cite{binnen} and HERA data \cite{h1parti}.

\section{Charmonium production}\label{sec:onium}

\begin{figure}[h]
\centerline{\psfig{figure=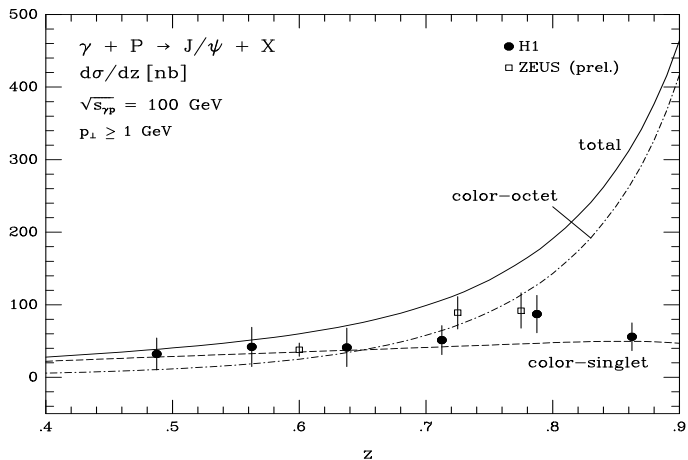,height=6.cm,width=9.cm}} 
\vspace{-.4cm}
\caption{{\em Fusion contribution to charmonium production at HERA.}} 
\label{fig:8}
\vspace{-.1cm}
\end{figure}

There has been new developments concerning the production of
hidden heavy flavor at the Tevatron where the usual model predictions,
based on the color singlet model, for prompt $\Psi$ production fall an
order of magnitude below the experimental results. The interest of HERA
lies in the fact, that many of the new parameters introduced in the
model can be tested independently. We consider here only the
non-diffractive mechanism, corresponding to a $\Psi$ inelasticity factor
of  $z=p.k_{_\Psi} / p.k_{\gamma}<.9$.
In the factorisation approach of ref. \cite{braatenbl} the cross section for the
production of a heavy quark bound state $H$ is
\bea
\vspace{-.3cm}
d\sigma(\gamma p \rightarrow H X) = \sum_{[{\bf n}]} 
d\hat{\sigma}(\gamma p \rightarrow Q\overline Q[{\bf n}] X) 
< {\cal O}^H[{\bf n}] >, 
\label{eq:psi}
\vspace{-.3cm}
\ena
where $d\hat{\sigma}$ is the cross section for producing a heavy
$Q\overline Q$ sytem in state $[\bf n]$ defined by its color [${\bf 1},
{\bf 8}$], its spin [$0, 1$] and orbital angular momentum. This term is
calculable perturbatively, the large scale beeing provided by the heavy
quark mass or eventually by the transverse momentum of the system. The
factor $< {\cal O}^H[\bf n] >$ describes the non-perturbative transition
from the state with quantum numbers $[\bf n]$ to quarkonium $H$.
Typically $< {\cal O}^\Psi[^3S_1,{\bf 1}] > = 1.16$ GeV$^3$ (from $\Psi$
leptonic decay) while the ${\bf 8}$ matrix elements 
$<{\cal O}^\Psi[^3S_1,{\bf 8}]> \sim$ 
$<{\cal O}^\Psi[^3S_0,{\bf 8}]> \sim  <{\cal O}^\Psi[^3P_J,{\bf 8}]>/m_c^2
 \sim 10^{-2}$ GeV$^3$ as determined from fits to the Tevatron
data \cite{chol}. The colour $\bf 8$ values are consistent with non-relativistic QCD which
predicts their suppression by powers of the velocity of the heavy quark
in the bound state \cite{lepage}. Although the $\bf 8$ matrix elements are small it may
happen that the perturbative  cross section $d{\hat \sigma}([\bf 8])$ is
large so that a non negligible contribution occurs. In the colour {\bf 1}
model the $\bf 8$ matrix elements are assumed to vanish. Many
processes  contribute to the cross section $d{\hat \sigma}( Q\overline
Q[\bf n])$: it is convenient to distinguish fusion processes which are
important at small $\pT$, but are suppressed by a factor $m_{_Q}^2 /
p_{_T}^2$, from fragmentation processes where the $Q \overline Q$ state
is found in the decay of a Q or gluon jet produced at large transverse
momentum. Furthermore, both processes come in the {\em direct} variety
if the photon couples to the hard sub-scattering, or the {\em resolved}
one if the photon structure function is involved. These mechanisms 
have different $z$ and $\pT$ dependences so that varying the kinematical
conditions they could be separated. \\
\begin{figure}[t]
\vspace{-.1cm}
\centerline{\psfig{figure=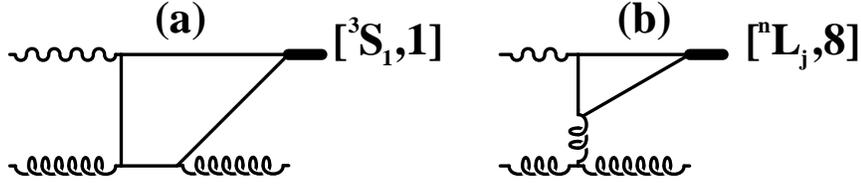}} 
\vspace{-.3cm}
\caption{{\em Some Feynman diagrams for charmonium production. 
({\bf a}): colour singlet production; ({\bf b}): diagram contributing only to 
colour octet production.}} 
\label{fig:7}
\vspace{-0.1cm}
\end{figure}
Consider first direct fusion processes which should be a good
approximation of cross sections at small transverse momentum ($p_{_T}
\le m_{_Q}$) or integrated over $p_{_T}$. The basic colour singlet
diagram is shown in fig. \ref{fig:7}a: the heavy $Q\bar Q$ pair is
produced by photon-gluon fusion and an extra gluon is necessarily
emitted so that a $Q\bar Q$ state in a {\bf 1} component can be
projected out. NLO corrections to this process have also been
calculated \cite{kraemer}. A similar diagram contributes also to the
production of a colour $\bf 8$ state together with the diagram of fig.
\ref{fig:7}b: because of the gluon exchange this diagram will be
enhanced near the kinematical region $z\sim 1$. The comparison of the
theory \cite{cacciakra} with HERA \cite{h1charm,zeuscharm}
data is shown in fig.  \ref{fig:8} where one sees the excellent
agreement between data and the colour $\bf 1$ NLO predictions while the
$\bf 8$  component seems to yield much too large a contribution at large
$z$ values. To claim quantitative disagreement is premature as the
non-perturbative $\bf 8$ matrix elements may have been overestimated in
the Tevatron analysis; furthermore, doubts have been raised about the
validity of the factorisation approach (velocity expansion) in the large
$z$ region \cite{beneke}. \\
\indent
The resolved fusion diagrams are easily obtained by substituting to the
photon a gluon, fragment of the photon: obviously this process dominates
at small $z$  where it is further found that the $\bf 8$ contribution
overwhelms the $\bf 1$ one by more than one order of magnitude: thus
the region at small $p_{_T}$, small $z$ or equivalently large rapidity
since $z \sim e^{-y_{_{\mbox {\cmr lab}}}}$, should be able to probe
the $\bf 8$ matrix elements.

At large $p_{_T}$, on the other hand, the production of  charmonium in
the fragments of a jet should be 
considered \cite{godbole,knikra}. 
In the factorisation approach, the fragmentation
funtion of parton $i$ into H takes the form:
\bea
D^H_i(z,M) = \sum_{[\bf n]} d_{i\rightarrow c\bar c[{\bf n}]}(z,M)
< {\cal O}^H[{\bf n}] >, \qquad i=c,g
\label{eq:frag}
\ena
\noindent where the functions $d_i$ are perturbatively calculated (all
scales involved are large) and the same non-perturbative matrix elements
as above appear. A detailed study at the NLO order \cite{knikra},
including both direct and resolved processes, has recently been
performed where it is found an overwhelming dominance of the
colour $\bf 8$ channels at HERA for large $y_{_{\mbox {\cmr lab}}}$
and $\pT$, the more so the larger the $\gamma p$ invariant mass. \\
The exclusive channel $\gamma p \rightarrow \Psi \gamma$ has also been
proposed \cite{cacciagk} as a test of the model as it is obviously
dominated at large $z$ by the $\bf 8$ component (see fig. \ref{fig:7}a
with the final state gluon changed into a photon).\\
Thus HERA appears a promising place to support or invalidate the
colour $\bf 8$ model if detailed measurements of specific channels
can be done over a wide rapidity range for $J\Psi$ production.

\section*{Acknowledgements}

I am grateful to M.~Erdmann for discussions and very helpful advice in
the  preparation of this talk. I also thank L.~Bourhis, I.~Butterworth,
M.~Cacciari, J.~Forshaw, J.Ph.~Guillet, B.W.~Harris, U.~Karshon,
B.~Kniehl, M.~Mangano, J.F.~Owens and E.~Pilon.
Support by the EEC program ``Human Capital and Mobility", Network
``Physics at High Energy Colliders", contract CHRX-CT93-0357 (DG 12
COMA), is greatefully acknowledged.


\end{document}